\documentclass[12pt,preprint]{aastex}
\usepackage{epsfig,graphicx} 

\usepackage{epsfig,emulateapj5,apjfonts,aastexug}

\def\pn{\par\noindent}

\def\gsimeq{\hbox{\raise0.5ex\hbox{$>\lower1.06ex\hbox{$\kern-1.07em{\sim}$}$}}} 
\def\lsimeq{\hbox{\raise0.5ex\hbox{$<\lower1.06ex\hbox{$\kern-1.07em{\sim}$}$}}} 

\newenvironment{onecolumnfigure}{
\def\@captype{figure}
\noindent\begin{minipage}{0.99\linewidth}\begin{center}}
{\end{center}\end{minipage}\smallskip}

\shorttitle{The cosmic history of X-ray Iron Line}
\shortauthors{Brusa, Gilli \& Comastri}

\begin{document}



\title{The iron line background}


\author{Marcella Brusa}
\affil{Max Planck Institut f\"ur extraterrestrische Physik, 
       Giessenbachstrasse 1, D--85748 Garching, Germany} 
\author{Roberto Gilli}
\affil{INAF -- Osservatorio Astrofisico di Arcetri, Largo E. Fermi 5, 
I--50125 Firenze, Italy}
\author{Andrea Comastri}
\affil{INAF --  Osservatorio Astronomico di Bologna, via Ranzani 1, 
I--40127 Bologna, Italy}


\begin{abstract}
\pn We investigate the presence of iron line emission among faint X-ray sources
identified in the 1Ms {\it Chandra} Deep Field South and in the 2Ms
{\it Chandra} Deep Field North. Individual source spectra are stacked
in seven redshift bins over the range $z=0.5-4$. We find that iron
line emission is an ubiquitous property of X--ray sources up to
z$\simeq$3. The measured line strengths are in good agreement with
those expected by simple pre--{\it Chandra} estimates based on X--ray
background synthesis models. The average rest frame equivalent width
of the iron line does not show significant changes with redshift.
\end{abstract}



\section{Introduction}
It has been pointed out already a decade ago (Matt \& Fabian 1994)
that prominent spectral features, and especially the $\sim6.4$ keV iron
K$\alpha$ emission line commonly found in Seyferts spectra (Nandra \&
Pounds 1994), may lead AGN synthesis models to predict a detectable
signature in the spectrum of the cosmic X--ray background (XRB) around
a few keV. \\
The integrated contribution of iron line emission in the
sources making the XRB has been quantitatively estimated by Gilli et
al. (1999), 
on the basis of a large database of observational results 
available in the pre-{\it Chandra} and XMM-{\it Newton} era.
The predicted shape of the integrated emission from
individual lines was computed following the prescriptions of the 
XRB synthesis model of Comastri et al. (1995) resulting in a rather broad 
bump extending from about 1 keV (depending on the maximum redshift assumed) 
up to 6.4 keV.  The expected intensity of such a bump above the 
XRB level is below the 5\% level over most of the energy range and reaches 
a maximum value of the order of 7\% around 2--3 keV, the exact value 
being dependent on the redshift ($z_{cut}$) at which the evolution 
of the luminosity function is stopped.  
Although such an estimate is model dependent, Gilli et
al. (1999) concluded that it can be safely regarded as an upper limit
to the total iron line contribution.  
\par\noindent 
In the last few years our knowledge of the XRB sources has been 
significantly improved by deep {\it Chandra} and XMM--{\it Newton} surveys. 
A fraction as large as 80--90\% of the XRB flux below 5--6 keV has 
been resolved into individual sources (Worsley et al.  2004).  Extensive
multiwavelength follow--up observations
have clearly established that
the X--ray source redshift distribution is peaked at $z\simeq$0.7--1
(e.g. Hasinger 2004, Gilli 2004) and that about 60\% of the XRB
originates at $z <$ 1 (Barger et al. 2003).  
In the XRB synthesis model worked out by Gandhi \& Fabian (2003) the
iron line contribution turned out to be maximum at about 3.8 keV
(corresponding to a typical redshift of 0.7) and the excess due to the
iron line is of the order of 3--4 \%.  \par Despite the increasing
number of independent observations of the XRB spectrum below 10 keV
with BeppoSAX (Vecchi et al.  1999), XMM--{\it Newton} (De Luca \&
Molendi 2004) and RXTE (Revnivtsev et al.  2004), yielding good signal to
noise ratio measurements of the extragalactic background, the 
accuracy reached so far is not such to detect the iron features at the
level expected by the model predictions outlined above.  \par An
alternative approach, devised to avoid the line smearing due to 
the large redshift range over which AGN spectra are summed
and the present uncertainties in the XRB spectrum, is to
search for iron features over appropriate redshift bins. The detection
of an iron line and the study of its intensity and profile would open
the possibility to investigate interesting issues such as the metal
abundances and the relative fraction of Compton thick sources (which
are known to have extremely strong iron lines) at high redshift, along with
relativistically broadened lines beyond the local Universe (Comastri,
Brusa, Civano 2004; Streblyanska et al. 2005). In the following we
will present and discuss the results obtained for a large sample of
spectroscopically identified sources in the X--ray surveys which
resolved most of the XRB: the 2 Ms pointing in the {\it Chandra} Deep
Field North (Alexander et al. 2003; hereinafter CDFN) and the 1 Ms
observation of the {\it Chandra} Deep Field South (Giacconi et
al. 2002; hereinafter CDFS).  Throughout the paper, the adopted values
for the Hubble constant and the cosmological parameters are H$_0$=70
km s$^{-1}$ Mpc$^{-1}$, $\Omega_{\Lambda}$=0.7, $\Omega_{\rm m}$=0.3.

\section{Data analysis}


\subsection{The sample} 

The optical spectroscopic and photometric redshifts of the X--ray
sources detected in the CDFN are presented by Barger et al. (2003),
while a list of spectroscopic redshifts of the sources detected in the
CDFS is reported in Szokoly et al. (2004; hereinafter
S04). Photometric redshifts based on good quality HST/ACS images and
deep ISAAC/VLT observations are available for all but four CDFS
sources without spectroscopic redshift (Zheng et al. 2004; hereinafter
Z04), leading to a virtually complete catalog of identified X--ray
sources. Here we have considered only those CDFS photo--z with a
quality flag greater than 0.5 (see Z04 for details) ensuring a
reliable redshift estimate. Moreover, given that both S04 and Z04
identifications are based on the X--ray source catalogue of Giacconi
et al. (2002), we have revised some optical/X--ray associations
according to the improved astrometry provided by Alexander et
al. (2003). Only sources with spectroscopic redshifts have been
included in the CDFN sample because no details are given on the
photo--z quality in Barger et al. (2003). The larger CDFS redshift
completeness allows us to obtain a counting statistic per redshift bin
similar to that of the CDFN despite the shorter exposure. \par
Individual source spectra have been stacked together in seven redshift
bins spanning the $z$=0.5--4 range (Table 1).  The choice of bin sizes
and distribution is driven by a trade--off between the number of
counts in each bin and the need to sample a corresponding observed
energy range narrow enough to detect the spectral feature, keeping at
the same time the instrumental response as uniform as possible.  For
these reasons, the redshift interval below $z$=0.5 is not considered;
moreover, the redshift interval 1.7 $<z<$ 2.5 is excluded because the
iron line is redshifted in the $\sim$ 1.8--2.4 keV energy range, which
encompasses the sharp drop in the effective area due to the
instrumental iridium edge.\\
The iron line signal is not expected to be significantly diluted by
non-AGN sources. Below z=2, the K$\alpha$ line falls in the hard band
($>2$ keV), where the detected sources are mostly AGN. At higher 
redshifts it is shifted to the soft band, but then only sources with
AGN luminosities can be detected. In order to avoid contamination by a
few individual bright sources which could dominate the stacked signal
we have excluded from each bin the brightest object, if its
contribution exceeds 50\% of the entire flux in the bin. The chosen
redshift bins, the expected position of the 6.4 keV line, the bin
width and the number of sources in each bin are listed in Table~1. The
final sample includes 171 sources in the CDFN and 181 in the CDFS,
spanning the luminosity range 
L$_{2-10 keV}=10^{41}-10^{45}$ erg s$^{-1}$ (see Table~1).
The small solid angle ($\sim$ 0.2 deg$^{-2}$) covered by the deep fields
does not allow us to sample the AGN population at higher luminosities.

\subsection{X--ray stacking}
The X--ray data for both the CDFN and CDFS observations have been retrieved 
from the public archive and processed with standard tools making 
use of the calibrations associated with the 
{\tt CIAO}\footnote{http://cxc.harvard.edu/ciao/} software (version 3.0). \\
A total of 20 and 11 pointings in the CDFN and CDFS, respectively, were 
summed together with the {\tt merge\_all}\footnote{http://cxc.harvard.edu/ciao3.0/threads/merge\_all/} 
script in two different merged event files. 
The Charge Transfer Inefficiency (CTI) and 
gain\footnote{http://hea-www.harvard.edu/alexey/acis/tgain} corrections
were also applied to each single pointing\footnote{We have also 
verified that applying the new gain correction tool 
released with the CALDB 2.28 version the results are not modified.}.
Spectra, response matrices and effective areas of the stacked spectrum 
in each redshift bin were extracted from the merged event files  
using the standard {\tt CIAO} tools developed to properly 
weight responses and effective area files for multiple extraction 
regions (e.g. {\tt mkwarf}; see Civano, Comastri \& Brusa 2005 for 
a detailed description of the procedure).
The target positions in each redshift bin span almost all the offaxis
angles, thus the extraction radius was varied from 4 to 16 arcsec depending
on the source brightness and off--axis position. 
Several background spectra were extracted with the same procedure from a stack
of regions nearby each source in each redshift bin and varied in size 
and shape. \\
Finally, the stacked spectra in each redshift bin of both the CDFN
and CDFS were summed together with standard {\tt FTOOLS} routines
({\tt mathpha, addarf, addrmf}) weighted for both the exposure
time and the number of counts. In the following we will consider only
the resulting CDFN+CDFS spectra.

\section{Spectral analysis}

Since we are interested in emission features over a well
defined energy range, source spectra have been extracted in the 1--6 keV
band, to minimize calibration uncertainties in the low energy response.
X--ray spectra were rebinned to have at least 20 counts per bin 
and fitted with {\tt XSPEC} (version 11.3.0; Arnaud 1996);
errors are reported at the 90$\%$
confidence level for one interesting parameter ($\Delta\chi^2$=2.71). \\
The continuum is parameterized with a single power--law with the
slope free to vary. A negative ($\tau$ = --0.17) edge at 
2.07 keV has been added to account for residual uncertainties in 
the effective area above 2 keV, as suggested by the CXC team (Vikhlinin et
al. 2005)\footnote{See also:
http://cxc.harvard.edu/ccw/proceedings/03\_proc/presentations/marshall2}.
With the exception of the highest redshift bin, a significant excess
above a power--law continuum is present over the energy range
6.4/(1+$z_{max}$) -- 6.4/(1+$z_{min}$) keV, where $z_{min}$ and
$z_{max}$ are the bin boundaries (see Fig.~1). We then added a
redshifted gaussian component (model {\tt po+zgauss}) with the width
free to vary, centroid energy fixed at 6.4 keV and redshift fixed at
the mean value of each bin ($\bar z$), and verified that in all
cases the fit improved, though with a different level of
significance. Table~1 lists the best fit parameters for each redshift
bin. 
The equivalent widths (EWs) quoted in Table~1 are measured in the
observed frame. To get the intrinsic, rest frame EWs a multiplicative
factor $1+\bar z$ must be applied\footnote{The superposition of 
sources at different redshift in each bin should return a slightly different
multiplicative factor, but the difference in the rest frame
EW obtained by using ($1+\bar z$) is negligible.}.

\subsection{Safety checks}

In order to check whether the results are dependent from the counting
statistics and/or the energy range over which the continuum underlying
the iron excess is measured, we have performed several safety checks.
\pn First, we have further excluded in each bin the two or three
brightest sources, if their contribution to the total counts was
larger than 60\%. Then, spectral fits were repeated in the 2.2--6 keV
energy range for sources at $z<2$. Finally, a redshifted absorption
component (assumed to be at $\bar z$ in each redshift bin) was added
to the continuum. In all cases, the spectral parameters were found in
agreement, within the statistical errors, with those reported in
Table~1. Furthermore, for the z$<2$ bins for which the statistics is
higher, we verified that, when the redshift of the line component is
left free to vary, the resulting best fit redshift is always 
almost coincident with that expected from a line at 6.4 keV.
\pn Extensive simulations have also been performed to
quantitatively assess the line broadening introduced by redshift
smearing. The spectra of the sources contributing to each bin were
modeled as a single power law plus a narrow unresolved ($\sigma$=0.1
keV) K$\alpha$ line. 
The actually observed sources redshift distribution within the bin 
is further assumed and
the individual simulated spectra are stacked together. With the exception
of the $z$=0.9--1.1 bin, all the measured widths are in agreement with
those resulting from the simulations, and therefore consistent with
being produced by the superposition of narrow features (but see also
the discussion).
\pn According to XRB synthesis models, a large fraction of the
sources contributing to each redshift bin are obscured by column 
densities in the range $\sim$10$^{22-24}$ cm$^{-2}$.
When obscured sources at different redshifts are stacked together, the
iron line underlying continuum is expected to be modified, with
respect to a single power law, by the most prominent absorption
features such as the low energy cut--off and the 7.1 keV iron edge.
To take this into account, we assumed a continuum shape as resulting
from the Gilli et al. (2001) model once the AGN luminosity function
is integrated in the narrow redshift bins adopted here. Although the
recomputed EW are lower than those reported in
Table 1 (by about 30-50\%), the line excess is still significant.
A more detailed investigation would require an extensive
analysis of the AGN synthesis models parameter space and is beyond the
purposes of this letter.
\begin{center}
\begin{onecolumnfigure}
\includegraphics[width=6.5cm]{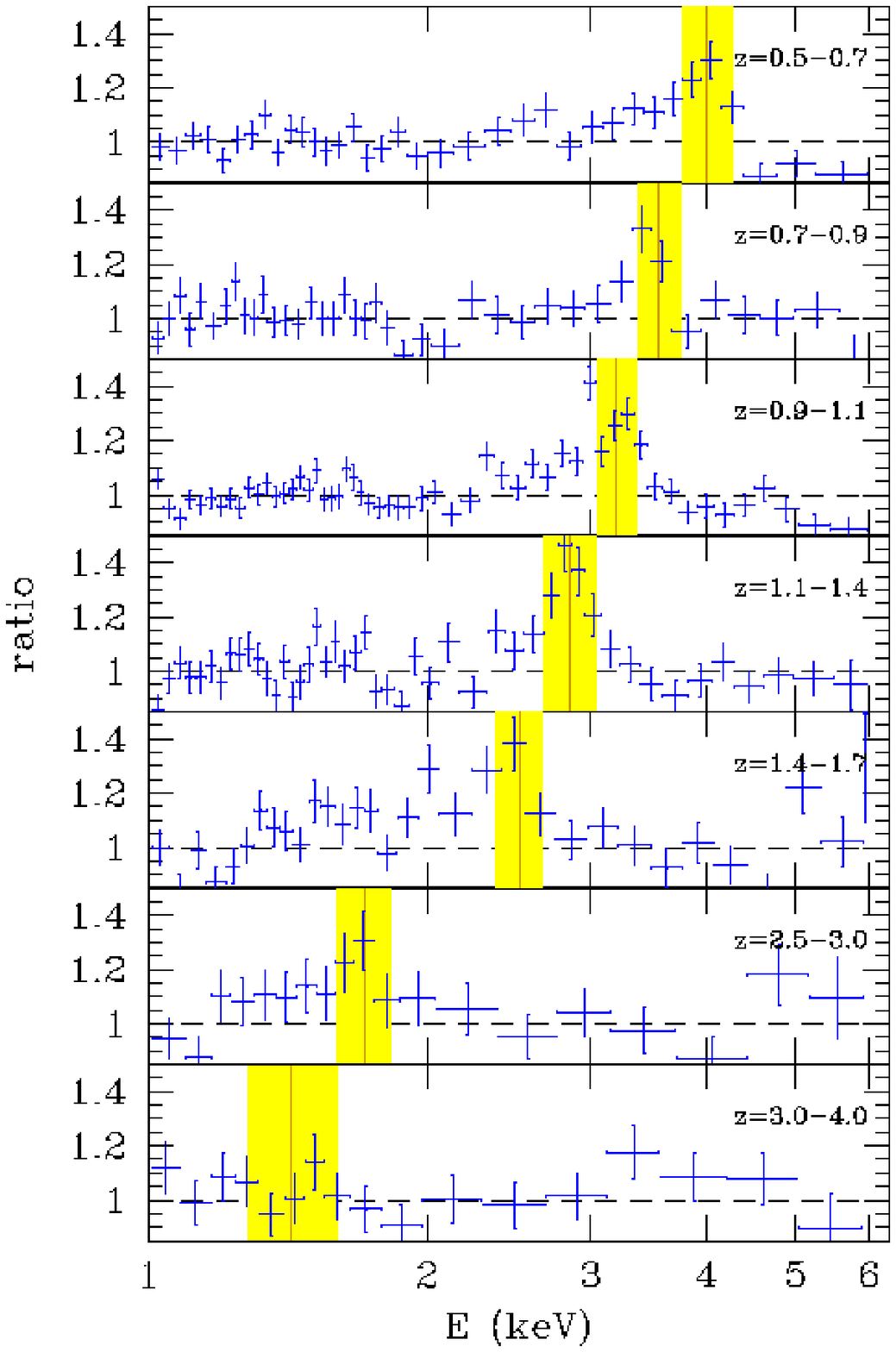}
\figcaption{Residuals of a
simple power--law fit to the source spectra in seven different
redshift bins as labeled. The vertical line in each panel is at the
expected position for the redshifted 6.4 keV Fe $K\alpha$ line while
the shaded region encompasses the bin width reported in Table~1 and
defined as $\Delta$E = 6.4/(1+$z_{max}$) -- 6.4/(1+$z_{min}$) keV.}
\label{plres}
\end{onecolumnfigure}
\end{center}

\section{Discussion}
\pn

The rest frame equivalent widths for the seven redshift bins 
are compared (Fig.~2) with those predicted by the XRB
synthesis model of Gilli et al. (2001, shaded area in Fig.~2).
The line EW distribution is modeled following Gilli et al. (1999, see their 
Table~4) and is a function of the absorption column density increasing
from about 250 eV for unobscured sources, to 400 eV for obscured
(log$N_H <$ 24) AGN, reaching about 2 keV for Compton thick sources.
These values are constant over the entire redshift range. 
The upper bound of the shaded region has been computed assuming 
EW=280 eV for unobscured sources, consistent with the average value 
measured by ASCA in the nearby Universe.
Recent XMM--{\it Newton} and {\it Chandra} results have questioned the
presence of broad iron lines in many unobscured-type 1 AGN, measuring
in turn a lower EW ($\sim 100-150$ eV on average) along with 
a decrease of the line EW towards high luminosities (e.g. Page et
al. 2004).  We have quantitatively estimated the effects of reducing
the line EW among unobscured sources by considering the most conservative
case of no iron emission at all (lower bound in Fig.~2). Since in our
model most of the iron signal is produced by obscured sources with
$22<$ log(N$_{rm H}$)$<24$ cm$^{-2}$, the overall EW are reduced by only
$15-20\%$.
\begin{center}
\begin{onecolumnfigure}
\includegraphics[width=6.5cm]{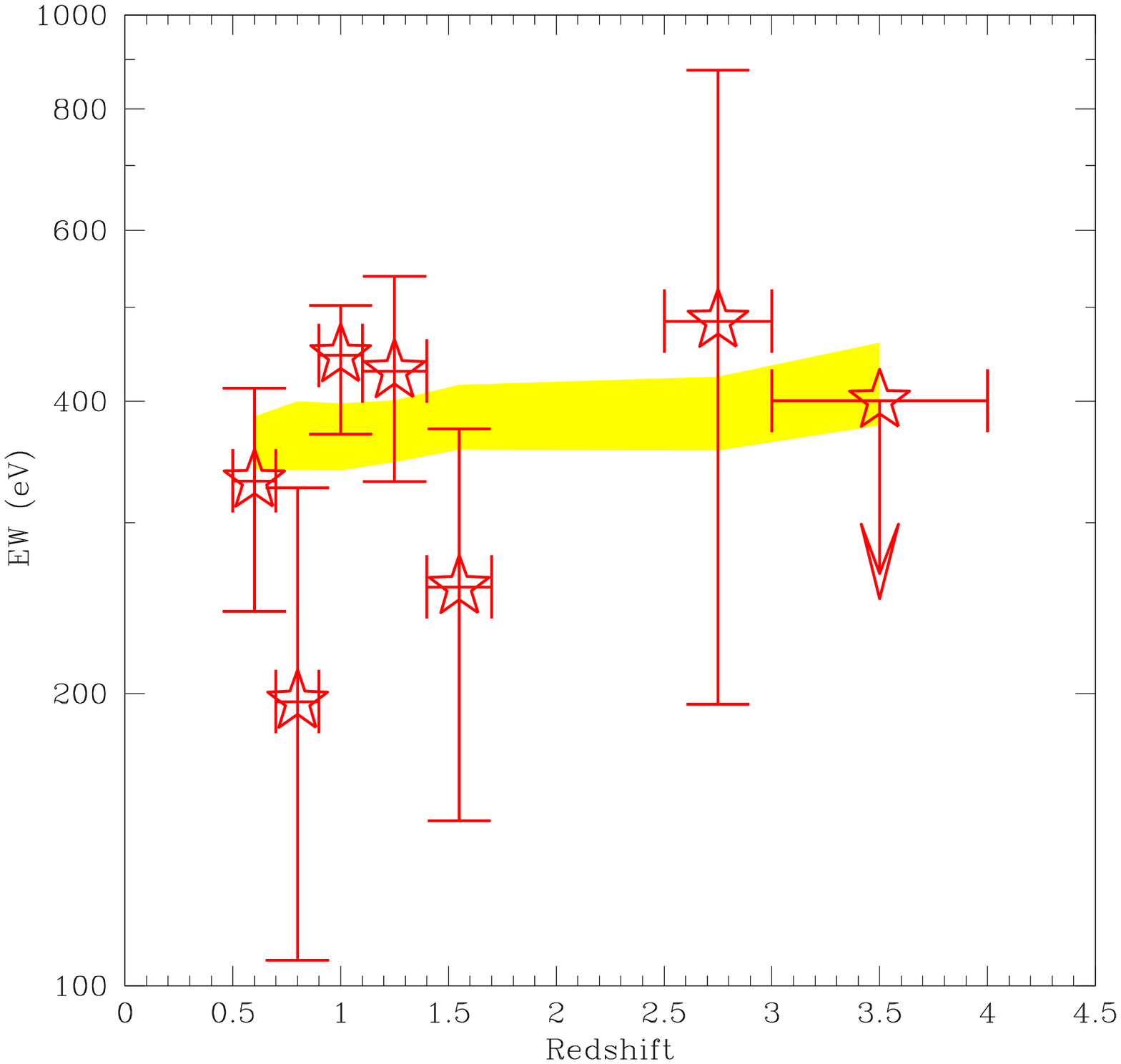}
\figcaption{Rest frame
equivalent widths as a function of redshift compared with the model
predictions (shaded strip; see text for details). }
\label{eqwz}
\end{onecolumnfigure}
\end{center}

The lack of any dependence of the average line 
intensity upon redshift 
up to $z\sim 3$ could be naively interpreted as a constant iron
abundance with redshift. The real case is likely to be much more
complicated, given the several factors affecting the detected
signal. 
As an example, the population 
of heavily obscured Compton thick sources (which have the strongest
iron  lines) might be hiding among the 
X--ray sources either unidentified or with a poorly constrained 
photo--z and thus not included in the present analysis.
However, according to our calculations (Gilli et al. 2001),
Compton Thick AGN are expected to provide only about 5--10\% of the
iron signal in each redshift bin. The presence of a population of
Compton Thick sources much larger than that assumed would be more
easily appreciated by the distortions introduced on the broad band
3--100 keV XRB spectrum rather than by a stronger line in the various
redshift bins.

A significant decrease of the line 
EW towards high luminosities has been reported  by 
Nandra et al. (1997) for a sample of type 1 AGN observed with ASCA.
The effect is more pronunced for luminosities higher than 
$10^{45}$ erg s$^{-1}$.  
Unfortunately, our sample is not well suited to investigate iron 
line properties at high luminosities  
due to the lack of sources luminous enough (see Table~1) 
to check the Nandra et al. (1997) findings. Moreover,
the stacked spectra are dominated by obscured sources rather
than type 1 AGN adding further uncertainties in the comparison.

Our findings are also consistent with those
reported by Streblyanska et al. (2005) from the analysis of the
rest--frame stacked spectrum of identified sources in the XMM--{\it
Newton} observation of the Lockman Hole ($\sim 400$ eV, 
when a gaussian profile is considered). 
The residuals leftward of the iron line which in our fits
are more prominent in the bins with the highest number of sources 
and counting statistic ($z$=0.5--0.7 and 0.9--1.1)
suggest the presence of a broad redshifted component, 
similar to that observed by Streblyanska et al. (2005).  
Since the main goal of the present
analysis is the detection of iron emission in the sources of the XRB
up to high redshifts, our approach is not designed to investigate in
detail the average line shape.  We recall, however, that the observed
EW and line profile are dependent on an accurate modeling of the
underlying continuum.  The superposition of absorbed spectra with
different redshifts has the effect of reducing the measured EW (see
previous Section) and furthermore may produce a spurious red wing.

\section{Summary}

We report the detection of iron emission up to $z\sim3$ in the X--ray
spectra of faint {\it Chandra} sources stacked into different redshift
bins. The measured EW are in agreement with a scenario in which the
lines are intrinsically narrow and their intensity does not change
significantly with redshift (and/or luminosity), which can be 
interpreted as a constant iron abundance as a function of
redshift. Extensive simulations and safety checks have been performed
in order to test the reliability of our results. Even taking into
account the effects of absorption features which modify the high
energy power law in obscured sources the detection of the emission
line remains significant. 
Despite their prominent iron lines, our approach does not allow to put 
tight constraints on the number density of Compton Thick sources, that 
suffer from extreme absorption in the {\it Chandra} band and can be detected 
only in small numbers compared to Compton thin AGN.
Although there might be hints for the presence of gravitationally 
redshifted broad line components, we caution that their intensity 
and profile significantly depends on the modeling of the underlying
continuum.

\acknowledgments
We gratefully thank Francesca Civano and Piero Ranalli for help in data 
reduction, Giorgio Matt, Gianni Zamorani, Cristian Vignali, Paolo Tozzi 
and Giancarlo Setti for useful discussions.
We acknowledge partial support from MIUR Cofin--03--02--23
and INAF/270/2003 grants.

\footnotesize
\scriptsize
\vspace{3.1cm}
\begin{center}
{\sc TABLE~1 \\ Redshift bins and number of sources}
\vskip 2pt
\end{center}
\begin{tabular}{cccccccrcccrr}
\hline \hline
z$^{(1)}$  & E($K\alpha$)$^{(2)}$ & Bin width$^{(3)}$ & 
\multicolumn{3}{c}{Number of sources$^{(4)}$} & $<L_{\rm X}>$$^{(5)}$ & Counts$^{(6)}$ & $\Gamma$$^{(7)}$ 
      & PL + zgauss$^{(8)}$  & $\Delta\chi^{2}$ $^{(9)}$ & $\sigma$$^{(10)}$ & EW$^{(11)}$ \\
min-max & keV & keV & CDFS  & CDFN & Total & $10^{42}$ erg s$^{-1}$ &        &
      & $\chi^{2}$/d.o.f.   &                  &  keV     & eV \\
\hline
0.5-0.7 & 4.00  &  0.50 & 45(7) & 39 & 84  & 1(0.1-140) & 16831      & 1.20$\pm0.03$
        & 284.2/289   &  44.2           &  0.30$^{+0.12}_{-0.09}$    
        & 207$_{-55}^{+51}$ \\
0.7-0.9 & 3.55  &  0.40 & 34(10) & 32 & 66 & 4(0.3-300) & 13240	& 1.44$\pm0.04$
        & 229.7/260   &  14.0           &  0.27$^{+0.78}_{-0.14}$    
        & 116$_{-58}^{+68}$\\
0.9-1.1 & 3.20 &   0.32 & 30(11) & 58 & 88 & 8(0.6-400) & 34958	& 1.47$\pm0.02$
        & 342.6/321   & 112.6           &  0.48$^{+0.14}_{-0.12}$    
        & 223$_{-36}^{+28}$\\
1.1-1.4 & 2.78 &   0.38 & 25(10) & 25 & 50 & 20(0.9-400) & 15629	& 1.45$\pm0.04$
        & 258.3/267   &  63.8          &  0.31$^{+0.15}_{-0.12}$    
        & 191$_{-44}^{+48}$\\
1.4-1.7 &  2.56  & 0.30 & 19(11) & 4  & 23 & 40(2-500) & 7841	& 1.36$\pm0.04$
        & 248.0/195   &  19.0          &  0.18$^{+0.18}_{-0.18}$    
        &  101$_{-43}^{+46}$\\
\hline
2.5-3.0 & 1.71   & 0.23 & 18(9) & 4  &  22  & 100(20-1000) & 4895	& 1.46$\pm0.07$
        & 164.5/141   & 10.0             & 1.15$^{+0.61}_{-0.40}$    
        & 129$_{-77}^{+105}$\\
3.0-4.0 &  1.42  & 0.32 & 9(3) & 9    & 18  & 200(36-2000) & 2919	& 1.30$\pm0.10$
        & 82.3/111     & 2.0              & $<1.94$
        & $<89$\\
\hline
\hline
\end{tabular}
\label{tabbin}
\pn Notes: 
$^{(1)}$Redshift range for the stacked spectra. 
$^{(2)}$Expected position of the Fe 6.4 keV line in the observed frame. 
$^{(3)}$Observed frame bin width, in keV, at the 6.4 keV position. 
$^{(4)}$Number of sources in the CDFS, CDFN and combined samples (in parenthesis for the CDFS sources with photometric redshifts).
$^{(5)}$Median 2-10 keV luminosity (in parenthesis the minimum and maximum 
luminosities)
$^{(6)}$Total net counts in the 1--6 keV band.
$^{(7)}$Best fit power--law spectral index and 90\% errors.
$^{(8)}$Statistical significance of the spectral fit for the {\tt po+zgauss} 
model. 
$^{(9)}$$\Delta\chi^{2}$ with respect to the single power--law model.
$^{(10,11)}$Best fit line width ($\sigma$) and observed frame intensity (EW) 
of the gaussian component and 90\% errors. 
\normalsize

\end{document}